\documentclass[twocolumn,pra,superscriptaddress,showpacs,tightenlines]{revtex4}
\usepackage{amssymb}
\usepackage{amsmath}
\usepackage{graphicx}
\usepackage{epsfig}
\usepackage{subfigure}
\usepackage{amsfonts}

\begin{document}

\title{Dynamic sensitivity of photon-dressed atomic ensemble with quantum criticality}
\author{Jin-Feng Huang}
\affiliation{Key Laboratory of Low-Dimensional Quantum Structures and Quantum Control of
Ministry of Education, and Department of Physics, Hunan Normal University,
Changsha 410081, China}
\author{Yong Li}
\affiliation{Department of Physics, The University of Hong Kong, Pokfulam Road, Hong
Kong, China}
\author{Jie-Qiao Liao}
\affiliation{Institute of Theoretical Physics, Chinese Academy of Sciences, Beijing,
100080, China}
\author{Le-Man Kuang}
\email{lmkuang@hunnu.edu.cn}
\affiliation{Key Laboratory of
Low-Dimensional Quantum Structures and Quantum Control of Ministry
of Education, and Department of Physics, Hunan Normal University,
Changsha 410081, China}
\author{C. P. Sun}
\email{suncp@itp.ac.cn} \homepage{http://www.itp.ac.cn/~suncp}
\affiliation{Institute of Theoretical Physics, Chinese Academy of
Sciences, Beijing, 100190, China}
\date{\today}

\begin{abstract}
We study the dynamic sensitivity of an atomic ensemble dressed by a
single-mode cavity field (called a photon-dressed atomic ensemble),
which is described by the Dicke model near the quantum critical
point. It is shown that when an extra atom in a pure initial state
passes through the cavity, the photon-dressed atomic ensemble will
experience a quantum phase transition, showing an explicit sudden
change in its dynamics characterized by the Loschmidt echo of this
quantum critical system. With such dynamic sensitivity, the Dicke
model can resemble to the cloud chamber for detecting a flying
particle by the enhanced trajectory due to the classical phase
transition.
\end{abstract}

 \maketitle


\section{\label{sec:1}Introduction}
The quantum phase transition (QPT)~\cite{Book} occurs at zero
temperature when the external parameters of some interacting
many-body systems change to reach the critical values. Generally, it
is associated with the ground state with energy level crossing and
symmetry breaking at the critical points. Recently, it was
discovered that, near the quantum critical point the QPT system
possesses the ultra-sensitivity in its dynamical
evolution~\cite{Quan2006}. This theoretical prediction has been
demonstrated by an NMR experiment~\cite{NMR}. Similar sensitivity
exists in some quantum
systems~\cite{Hepp,Emary2003,Zanardi,Fazio,Zhang,Wang} possessing
QPT.

In this paper, we study the dynamic sensitivity of an atomic
ensemble in a cavity with a single-mode electromagnetic field
(called a photon-dressed atomic ensemble), which is described by the
Dicke model~\cite{Dicke}. We assume the atoms in the Dicke model are
resonant with the cavity field. When an extra two-level atom in
large detuning goes through the cavity field, the frequency of
cavity field will be shifted effectively according to the Stark
effect so that the photon-dressed atomic ensemble near the QPT will
be forced into its critical point. In this situation the dynamic
evolution of the Dicke model becomes too sensitive in response to
the passage of the extra atom.

Here, this dynamic sensitivity is characterized by the Loschmidt
echo (LE)~\cite{LE}, which is intrinsically defined by the structure
of the photon-dressed atomic ensemble. For a short time
approximation, we prove that the LE is just an exponential function
of the photon number variance in the photon-dressed atomic ensemble.
This finding means that the LE can be experimentally measured by
detecting the photon correlation. Its sudden change may imply the
passage of an extra atom through the cavity. With this
reorganization we will demonstrate that such quantum sensitivity in
the Dicke model is very similar to the classical sensitivity of the
cloud chamber for detecting a flying particle, which is
characterized by the macroscopically observable trajectories
enhanced by the classical phase transitions.
\begin{center}
\begin{figure}[h]
\includegraphics[bb=0 2 339 293, width=7 cm]{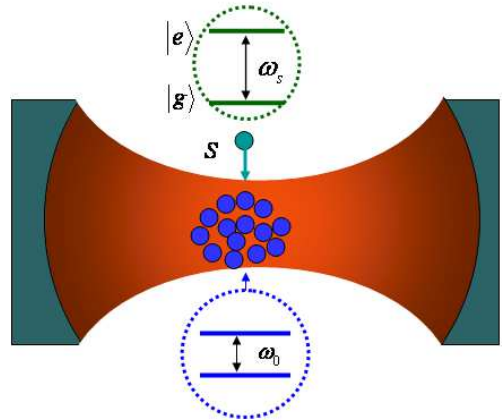}
\caption{(Color online) Schematic of a cavity field coupled with an
atomic gas consisting of $N$ two-level atoms. An extra detected
two-level atom $S$ is injected into cavity field.} \label{fig1}
\end{figure}
\end{center}

This paper is organized as follows. In Sec.~\ref{sec:2}, we describe
the setup of the quantum critical model based on the Dicke model.
The effective Hamiltonian is given in terms of the collective
excitation of the atomic ensemble. Then the analytic calculation of
LE (or the decoherence of the extra atom) is carried out in
Sec.~\ref{sec:3} for the normal and super-radiant phases,
respectively, by short-time approximation. In the following
Sec.~\ref{sec:4} we plot some figures to explicitly show the
sensitive properties of the LE. In Sec.~\ref{sec:5}, we address the
similarity between the dynamic sensitivity of the photon-dressed
atomic ensemble induced by an extra atom and the classical cloud
chamber. Finally, we draw our conclusion in Sec.~\ref{sec:6}. The
detailed coefficients for Bogoliubov transformation in Sec. IV are
given in the Appendix.
\section{\label{sec:2}Model and Hamiltonian}
As showed in Fig.~\ref{fig1}, we consider an atomic ensemble
confined in a gas cell coupled with a single-mode cavity field of
frequency $\omega $, which is described by the annihilation
(creation) operator $a$ ($a^{\dag }$). We use the Pauli matrices
$\sigma _{z}^{(j)}=|e\rangle _{jj}\langle e|-|g\rangle _{jj}\langle
g|$, $\sigma _{+}^{(j)}=|e\rangle _{jj}\langle g|$, and $\sigma
_{-}^{(j)}=|g\rangle _{jj}\langle e|$ to describe the atomic
transition of the $j$th atom with energy level spacing $\omega
_{0}$, where $\left\vert e\right\rangle _{j}$ and $\left\vert
g\right\rangle _{j}$ are the excited and ground states of the $j$th
atom, respectively. The system of the atomic ensemble coupled with
the single-mode cavity field is described by the Dicke model
(hereafter, we take $\hbar =1$),
\begin{equation}
H_{0}=\omega a^{\dag }a+\sum_{j=1}^{N}\left[ \frac{1}{2}\omega _{0}\sigma
_{z}^{(j)}+g_{0}(a^{\dag }+a)\left( \sigma _{-}^{(j)}+\sigma
_{+}^{(j)}\right) \right] .  \label{H-0}
\end{equation}
Here, for small-dimension atomic gas~\cite{Emary2003}, we have
assumed that all the atoms locate near the origin point and interact
with the cavity field with the identical coupling strength $g_{0}$.

An extra two-level atom $S$ with transition operators $\sigma _{z}$,
$\sigma _{+}$, and $\sigma _{-}$ couples to the original single-mode
cavity field with Hamiltonian
\begin{equation}
H_{I}=\frac{1}{2}\omega _{s}\sigma _{z}+g_{s}(a^{\dag }\sigma _{-}+a\sigma
_{+}),
\end{equation}
where we have made a rotating wave approximation. Similarly, $\omega
_{s}$ is the transition frequency between the ground state
$\left\vert g\right\rangle $ and excited state $\left\vert
e\right\rangle $ of the atom $S$; $g_{s}$ is the corresponding
coupling strength.

It has been shown that the QPT will occur in the system described by
Dicke Hamiltonian~(\ref{H-0})~\cite{Emary2003}, since it keeps
Hermitian only for a small coupling strength $g_{0}$. But it is only
a model to display QPT in quantum optical system. Actually it could
not happen for the realistic atomic, molecular, and optical (AMO)
system if the unreasonably ignored two-photon term $A^{2}$ is
included~\cite{Rzazewski1975}. To focus on our main idea in the
work, we only regard the Dicke system as a simplified model. We
would like to point out that many authors have recognized this
problem, but there still exist many explorations by using this
simplified model~\cite{Dicke model}.

If the atom $S$ is far-off-resonant with the cavity field, that is,
the detuning $\Delta _{s}$ ($\equiv\omega _{s}-\omega $) is much
larger than the corresponding coupling strength $g_{s}$, i.e.,
$|\Delta _{s}|$ $\gg $ $g_{s}$, then one can use the so-called
Fr\"{o}hlich-Nakajima transformation~\cite{Frohlich,Nakajama} (or
other elimination methods) to obtain the effective total Hamiltonian
\begin{eqnarray}
H_{\mathrm{eff}}&=&(\omega +\tilde{\delta}\sigma _{z})a^{\dag
}a+\frac{1}{2}(\omega _{s}+\tilde{\delta})\sigma_{z}+\frac{\omega
_{0}}{2}\sum_{j=1}^{N}\sigma _{z}^{(j)}\notag\label{e2} \\
&&+\frac{g}{\sqrt{N}}\sum_{j=1}^{N}( a^{\dag }+a)\left( \sigma
_{-}^{\left( j\right) }+\sigma _{+}^{\left( j\right)}\right),
\end{eqnarray}
where $\tilde{\delta}\equiv g_{s}^{2}/\Delta _{s}$ and $g\equiv
g_{0}\sqrt{N} $. We note that the Fr\"{o}hlich-Nakajima transformation
is equivalent to the approach based on the adiabatical elimination.

The Hilbert space of $N$ two-level atoms is spanned by $2^{N}$ basis
states. In the current case all the atoms have the same free
frequencies and coupling constants with the cavity field, we can
consider these atoms being identical. Then the Hilbert space is
reduced into a subspace of $(2N+1)$ dimension. In this subspace,
Hamiltonian~(\ref{e2}) is simplified by introducing the collective
atomic operators
\begin{equation}
J_{\pm }=\sum_{j=1}^{N}\sigma _{\pm }^{\left( j\right) },\hspace{0.5
cm}J_{z}=\frac{1}{2}\sum_{j=1}^{N}\sigma _{z}^{\left( j\right)},
\end{equation}
which obey the following angular momentum commutation relations,
\begin{equation}
\lbrack J_{z},J_{\pm }]=\pm J_{\pm },\hspace{0.5cm}[J_{+},J_{-}]=2J_{z}.
\end{equation}
The collective atomic operator $J_{z}$ denotes the collective
population of the atomic gas and $J_{\pm }$ represents the
collective transitions.

In terms of the above angular momentum operators, Hamiltonian
(\ref{e2}) is written as
\begin{eqnarray}
H_{\mathrm{eff}}&=&(\omega +\tilde{\delta}\sigma _{z})a^{\dag
}a+\frac{1}{2}
(\omega _{s}+\tilde{\delta})\sigma _{z}  \notag \\
&&+\omega _{0}J_{z}+\frac{g}{\sqrt{N}}( a^{\dag }+a) \left(
J_{+}+J_{-}\right),
\end{eqnarray}
which is further reduced to
\begin{eqnarray}
H_{\mathrm{eff}} &=&(\omega +\tilde{\delta}\sigma _{z})a^{\dag }a+\omega
_{0}b^{\dag }b+\frac{1}{2}(\omega _{s}+\tilde{\delta})\sigma _{z}  \notag \\
&&+g(a^{\dag }+a)\left( b^{\dag }\sqrt{1-b^{\dag }b/N}+h.c.\right)
\label{H-eff}
\end{eqnarray}
(up to constant terms) through making use of the
Holstein-Primakoff~\cite{HP} transformation, which represents the
angular momentum operators in terms of a single bosonic mode as
follows:
\begin{eqnarray}
J_{+} &=&b^{\dag }\sqrt{N-b^{\dag }b},\notag\\
J_{-} &=&\sqrt{N-b^{\dag }b}b,\notag\\
J_{z} &=&b^{\dag }b-\frac{1}{2}N.
\end{eqnarray}

To see more explicitly the dynamic sensitivity of the photon-dressed
atomic ensemble in response to the extra atom, corresponding to
different state of the extra atom, the effective Hamiltonian in
Eq.~(\ref{H-eff}) reads
\begin{equation}
H_{\mathrm{eff}}=\left\vert g\right\rangle \left\langle g\right\vert
\otimes H_{g}+\left\vert e\right\rangle \left\langle e\right\vert
\otimes H_{e}
\end{equation}
with
\begin{eqnarray}
H_{g} &=&\omega _{g}a^{\dag }a+\omega _{0}b^{\dag }b+g(a^{\dag }+a)  \notag
\\
&&\times \left( b^{\dag }\sqrt{1-b^{\dag }b/N}+h.c.\right),  \label{Hg} \\
H_{e} &=&\omega _{e}a^{\dag }a+\omega _{0}b^{\dag }b+g(a^{\dag }+a)  \notag
\\
&&\times \left( b^{\dag }\sqrt{1-b^{\dag }b/N}+h.c.\right),
\label{He}
\end{eqnarray}
where $\omega _{e}=\omega +\tilde{\delta}$ and $\omega _{g}=\omega
-\tilde{\delta}$. Note that in the derivation of the above
Hamiltonians~(\ref{Hg}) and~(\ref{He}), we have discarded some constant
terms.

\section{\label{sec:3}Quantum critical Effect}

Before the extra atom $S$ is sent into the cavity, the
photon-dressed atomic ensemble (including the cavity field and the
atomic gas) is described by the Dicke Hamiltonian
\begin{eqnarray}
H_{G} &=&\omega a^{\dag }a+\omega _{0}b^{\dag }b+g(a^{\dag }+a)  \notag \\
&&\times \left( b^{\dag }\sqrt{1-b^{\dag }b/N}+h.c.\right).
\label{HG}
\end{eqnarray}
Comparing Eqs.~(\ref{Hg}) and (\ref{He}) with Eq.~(\ref{HG}), we
find, as a result of the injection of the atom $S$, only the
frequency of the optical field changes by a small shift
$\tilde{\delta}$ in the dynamic evolution of the photon-dressed
atomic ensemble.

The photon-dressed atomic ensemble is initially prepared in the
ground state $\left\vert G\right\rangle $ of Hamiltonian (\ref{HG})
and the extra atom $S$ in a superposed state $\alpha \left\vert
g\right\rangle +\beta \left\vert e\right\rangle $, where the
normalization condition requires $ |\alpha |^{2}+|\beta |^{2}=1$.
When the extra atom $S$ interacts dispersively with the cavity
field, the total system is governed by Hamiltonians~(\ref{Hg}) and
(\ref{He}) corresponding to the extra atom $S$ in states $\left\vert
g\right\rangle $ and $\left\vert e\right\rangle $, respectively.
Then at time $t$ the state of the total system becomes an
entanglement one,
\begin{eqnarray}
\vert \Psi (t)\rangle  &=&e^{-iH_{\mathrm{eff}}t}(\alpha \vert
g\rangle +\beta \vert e\rangle )\otimes
\vert G\rangle   \notag \\
&=&\alpha \vert g\rangle \otimes e^{-iH_{g}t}\vert G\rangle +\beta
\vert e\rangle \otimes e^{-iH_{e}t}\vert G\rangle \notag \\
&\equiv &\alpha\vert g\rangle \otimes\left\vert
G_{g}(t)\right\rangle +\beta\vert e\rangle\otimes\left\vert
G_{e}(t)\right\rangle , \label{psi}
\end{eqnarray}
where we have defined
\begin{eqnarray}
\left\vert G_{g}(t)\right\rangle\equiv e^{-iH_{g}t}\vert G\rangle,
\hspace{0.5 cm} \left\vert G_{e}(t)\right\rangle\equiv
e^{-iH_{e}t}\vert G\rangle.
\end{eqnarray}
 The generation of the above entanglement is due to
the conditional dynamics of the total system. This is to say,
corresponding to the detected atom prepared in states $\left\vert
g\right\rangle $ and $\left\vert e\right\rangle $, the evolution of
the photon-dressed atomic ensemble will be governed by the
Hamiltonians $H_{g}$ and $H_{e}$, respectively. The central task of
this paper is to show that the dynamic of the photon-dressed atomic
ensemble is sensitive to the state of the extra atom. When the
photon-dressed atomic ensemble stays in the vicinity of the QPT, the
effect of QPT must impose on the state of the extra atom with some
enhancement fashion, like the results in Ref.~\cite{Quan2006}. This
motivates us to study the quantum decoherence of the extra atom near
the critical point of the photon-dressed atomic ensemble, which can
also reflect the dynamic sensitivity of the photon-dressed atomic
ensemble.

By tracing over the degree of freedom of the photon-dressed atomic
ensemble in evolution state~(\ref{psi}), the reduced density matrix
$\rho _{s}(t)=\mathtt{Tr}_{a,b}\{\left\vert \Psi
\left(t\right)\right\rangle \left\langle\Psi\left(t\right)
\right\vert\}$ of the detected atom is obtained as
\begin{equation}
\rho _{s}(t)=|\alpha |^{2}\left\vert g\right\rangle \left\langle
g\right\vert +|\beta |^{2}\left\vert e\right\rangle \left\langle
e\right\vert +(D\alpha ^{\ast }\beta \left\vert e\right\rangle
\left\langle g\right\vert +h.c.),
\end{equation}
where we have introduced the decoherence factor
\begin{equation}
D(t)=\left\langle G\right\vert \exp ( iH_{g}t) \exp \left(
-iH_{e}t\right) \left\vert G\right\rangle .
\end{equation}
Alternatively, we can investigate the decoherence of the extra atom
by examining the so-called LE
\begin{equation}
L(t)\equiv \left\vert D(t)\right\vert ^{2}
\end{equation}
defined for the dynamic sensitivity of the photon-dressed atomic
ensemble. For a short time $t$, the LE can be approximated as
\begin{equation}
L(t)\approx \left\vert \left\langle G\right\vert
e^{-2i\tilde{\delta} ta^{\dag }a}\left\vert G\right\rangle
\right\vert ^{2}.\label{L(t)}
\end{equation}
The straightforward calculation can give
\begin{equation}
L(t)\approx \exp \left(-4\gamma \tilde{\delta}^{2}t^{2}\right).
\label{calculate-b}
\end{equation}
Here, we have introduced the photon number variance
\begin{equation}
\gamma\equiv\left\langle\left( a^{\dag}a\right)^{2}\right\rangle
-\left\langle a^{\dag}a\right\rangle^{2},\label{variance}
\end{equation}
and the average $\langle\cdot \rangle$ is taken for the ground state
$\left\vert G\right\rangle$.

We point out that, up to the second order of time $t$, the decay
rate of the LE depends not only on $t^{2}$, but also on the photon
number variance $\gamma $. It is well known that the photon-dressed
atomic ensemble described by Dicke Hamiltonian~(\ref{HG}) transits
from the normal phase to the super-radiant one with the increase in
the parameter $g$ from that less than the critical value
$g_{c}=\sqrt{\omega \omega _{0}}/2$ to that larger than $g_{c}$.
Going across the phase transition point, the ground state of the
photon-dressed atomic ensemble experiences a complex change. We can
predict that the photon number variance $\gamma $ of the ground
state will exhibit some special features at the critical point.

According to Eq.~(\ref{psi}), we can imagine that the quantum
criticality of the photon-dressed atomic ensemble can display which
single state $\vert g\rangle $ or $\vert e\rangle $ that the extra
atom stays. When $L(t)$ approaches zero, the photon-dressed atomic
ensemble is forced into two orthogonal states $\vert
G_{g}(t)\rangle$ and $\vert G_{e}(t)\rangle$, and thus it behaves as
a measurement apparatus to detect the state of the extra atom. In
this case, its measurement on the atom will induce the decoherence
of the extra atom.

In what follows, we will calculate the photon number variance
$\gamma $ of the photon-dressed atomic ensemble in two different
phases, that is, the normal phase and the super-radiant phase.

\subsection{Dynamic sensitivity in normal phase}

In this subsection, we explicitly calculate $\gamma $ to investigate
the properties of the LE when the photon-dressed atomic ensemble is
within the normal phase. In the case of low excitations at
thermodynamic limit $N\rightarrow \infty $, Hamiltonian~(\ref{HG})
becomes
\begin{equation}
H_{G}=\omega a^{\dag }a+\omega _{0}b^{\dag }b+g(a^{\dag }+a)(b^{\dag }+b)
\label{Hnormal}
\end{equation}
for $\sqrt{1-b^{\dag }b/N}\approx 1$, which is typical to describe
two-coupled harmonic oscillators. It is well known that
Hamiltonian~(\ref{Hnormal}) becomes non-Hermitian in the over-strong
coupling region $g>g_{c}$, namely, the Hamiltonian possesses
imaginary eigenvalues~\cite{wagner}. This means effective
Hamiltonian~(\ref{Hnormal}) is ill-defined for $g>g_{c}$. Therefore,
we now restrict the Hamiltonian within the so-called normal phase
region $g<g_{c}$. Correspondingly, this limited
Hamiltonian~(\ref{Hnormal}) describes the normal phase of the Dicke
model.

In the normal phase, Hamiltonian~(\ref{Hnormal}) can be diagonalized
as
\begin{equation}
H_{G}=\omega _{A}A^{\dag }A+\omega _{B}B^{\dag }B  \label{e19}
\end{equation}
by introducing the polariton operators $A$ ($A^{\dag }$) and $B$
($B^{\dag }$), which depict the mixed bosonic fields of photons and
collective atomic excitations. The eigen-frequencies of the
polaritons $A$ and $B$ are
\begin{eqnarray}
\omega _{A}^{2} &=&\frac{1}{2}(\omega _{0}^{2}+\omega ^{2})-\frac{1}{2}\sqrt{%
(\omega _{0}^{2}-\omega ^{2})^{2}+16g^{2}\omega _{0}\omega }, \\
\omega _{B}^{2} &=&\frac{1}{2}(\omega _{0}^{2}+\omega ^{2})+\frac{1}{2}\sqrt{%
(\omega _{0}^{2}-\omega ^{2})^{2}+16g^{2}\omega _{0}\omega }.
\end{eqnarray}
It is straightforward to see that $\omega _{A}^{2}<0$ when
$g>g_{c}\equiv\sqrt{\omega\omega_{0}}/2$. That is, the
eigen-frequency $\omega_{A}$ of mode $A$ becomes a complex number,
which means Hamiltonian~(\ref{e19}) will be non-Hermitian in the
coupling region of $g>g_{c}$.

The relations between the operators \{$a$, $b$, $a^{\dag }$, $b^{\dag }$\}
and \{$A$, $B$, $A^{\dag },B^{\dag }$\} are given by
\begin{eqnarray}
a^{\dag } &=&f_{1}A^{\dag }+f_{2}A+f_{3}B^{\dag }+f_{4}B,  \notag \\
b^{\dag } &=&h_{1}A^{\dag }+h_{2}A+h_{3}B^{\dag }+h_{4}B,
\label{a-AB}
\end{eqnarray}
where the concrete forms of coefficients $f_{i}$ and $h_{i}$
($i=1,2,3,4$) have been given by Ref.~\cite{Emary2003}. Here we only
give the detailed forms of $f_{i}$ in the Appendix.

From Eq.~(\ref{e19}), we can see that the ground state of the
photon-dressed atomic ensemble in the polariton representation is
$|G\rangle =|0\rangle _{A}\otimes|0\rangle _{B}\equiv |00\rangle
_{AB}$. Making use of Eqs.~(\ref{variance}) and (\ref{a-AB}), we can
obtain the photon number variance
\begin{equation}
\gamma =2f_{1}^{2}f_{2}^{2}+2f_{3}^{2}f_{4}^{2}+\left(
f_{1}f_{4}+f_{2}f_{3}\right) ^{2}.  \label{LEnormal}
\end{equation}
In the normal phase, all the coefficients $f_{i}$ ($i=1,2,3,4$) are real,
then the photon variance is a positive number, which implies the coherence
of the extra atom will vanish with time.

We have mentioned that Hamiltonian~(\ref{Hnormal}) of two-coupled
harmonic oscillators can not work well in the over-strong coupling
region ($g>g_{c}$). This is because the approximation
$\sqrt{1-b^{\dag }b/N}\approx 1$ for the original one
[Eq.~(\ref{HG})] can not make sense in this region. Thus, we need to
consider a different approximation for Eq.~(\ref{HG}) when
$g>g_{c}$.

\subsection{Dynamic sensitivity in super-radiant phase}

Physically, when the atom-light coupling becomes stronger and
stronger, the coupled system will acquire a macroscopic excitations
of atomic ensemble. And then the system enters into a super-radiant
phase when $g>g_{c}$. In this situation, the low-excitation
approximation is no longer valid. We can use the coherent state
$\left\vert \beta \right\rangle$ of the collective atomic operator
$b$ to depict these kinds of macroscopic excitations~\cite{Hepp}. To
achieve the effective Hamiltonian over such background of
macroscopic excitations, we need to do the
displacement~\cite{Hepp,Emary2003}
\begin{equation}
b^{\dag }\rightarrow b^{\prime \dag }-\sqrt{\beta }
\end{equation}
(or alternatively, $b^{\dag }\rightarrow b^{\prime \dag }+\sqrt{\beta }$).
Correspondingly, we also displace the optical field by
\begin{equation}
a^{\dag }\rightarrow a^{\prime \dag }+\sqrt{\alpha }
\end{equation}
(or alternatively, $a^{\dag }\rightarrow a^{\prime \dag
}-\sqrt{\alpha }$). Here $a^{\prime \dag }$ and $b^{\prime \dag }$
describe quantum fluctuations about the semiclassical steady
state~\cite{Carmichael}; elsewhere, $\sqrt{\alpha }$ and
$\sqrt{\beta }$ describe the macroscopic mean fields above $g_{c}$
in the order of $O(\sqrt{N})$ \cite{Emary2003}. Then
Hamiltonian~(\ref{HG}) becomes
\begin{eqnarray}
H_{G} &=&\omega _{0}\left[ b^{\prime \dag }b^{\prime }-\sqrt{\beta }
(b^{\prime \dag }+b^{\prime })+\beta \right]  \notag  \label{HG-super1} \\
&&+\omega \left[ a^{\prime \dag }a^{\prime }+\sqrt{\alpha }(a^{\prime \dag
}+a)+\alpha \right]  \notag \\
&&+g\sqrt{\frac{k}{N}}\left( a^{\prime \dag }+a^{\prime
}+2\sqrt{\alpha}
\right)  \notag \\
&&\times \left(b^{\prime \dag }\sqrt{\xi }+\sqrt{\xi }b^{\prime
}-2\sqrt{ \beta }\sqrt{\xi }\right),
\end{eqnarray}
where
\begin{equation}
\sqrt{\xi }=\sqrt{1-[d^{\dag }d-\sqrt{\beta }(d^{\dag }+d)]/(N-\beta )}
\notag
\end{equation}
is introduced. In the thermodynamic limit $N\rightarrow \infty$, for
Eq.~(\ref{HG-super1}), we follow Emary and Brandes~\cite{Emary2003}:
expand the square root $\sqrt{\xi}$ and keep terms up to the order of
$N^{0}$ in the Hamiltonian. Then through choosing the appropriate
displacements
\begin{equation*}
\sqrt{\alpha }=\frac{g}{\omega }\sqrt{N(1-\mu ^{2})},\hspace{0.5cm}\sqrt{%
\beta }=\sqrt{\frac{N}{2}(1-\mu )}
\end{equation*}%
with $\mu =\omega \omega _{0}/4g^{2}$, we can diagonalize
Hamiltonian (\ref{HG-super1}) as
\begin{equation}
H_{G}=\omega _{A}^{\prime }A^{\prime \dag }A^{\prime }+\omega
_{B}^{\prime }B^{\prime \dag }B^{\prime }\label{e27}
\end{equation}
by the Bogoliubov transformation
\begin{eqnarray}
a^{\prime \dag } &=&f_{1}^{\prime }A^{\prime \dag }+f_{2}^{\prime }A^{\prime
}+f_{3}^{\prime }B^{\prime \dag }+f_{4}^{\prime }B^{\prime },
\label{coefficientsinsrp} \notag\\
b^{\prime \dag } &=&h_{1}^{\prime }A^{\prime \dag }+h_{2}^{\prime
}A^{\prime }+h_{3}^{\prime }B^{\prime \dag }+h_{4}^{\prime
}B^{\prime },
\end{eqnarray}
where the coefficients $f_{i}^{\prime }$ and $h_{i}^{\prime }$
($i=1,2,3,4$) have been given in Ref.~\cite{Emary2003}. Here we only
give the detailed forms of $f_{i}^{\prime }$ in the Appendix.

The eigen-frequencies $\omega _{A}^{\prime }$ and $\omega _{B}^{\prime
}$ of the polaritons described by the operators $A^{\prime }$ and
$B^{\prime }$ are given by
\begin{eqnarray}
\omega _{A}^{^{\prime }2} &=&\frac{1}{2}\left[\frac{\omega _{0}^{2}}{\mu ^{2}%
}+\omega ^{2}-\sqrt{\left( \frac{\omega _{0}^{2}}{\mu ^{2}}-\omega
^{2}\right) ^{2}+4\omega ^{2}\omega _{0}^{2}}\right],  \label{omegap} \\
\omega _{B}^{^{\prime }2} &=&\frac{1}{2}\left[\frac{\omega _{0}^{2}}{\mu ^{2}%
}+\omega ^{2}+\sqrt{\left( \frac{\omega _{0}^{2}}{\mu ^{2}}-\omega
^{2}\right) ^{2}+4\omega ^{2}\omega _{0}^{2}}\right].
\end{eqnarray}
It is known that if the coupling strength $g$ exceeds the critical
value $g_{c}$, both the above eigen-frequencies are real, but not in
the region of $g<g_{c}$. Namely, when $g>g_{c}$, Hamiltonian
(\ref{e27}) is Hermitian.

In the super-radiant phase, the ground state $\vert G\rangle =\vert
00\rangle _{A^{\prime }B^{\prime }}$ satisfies $A^{\prime }\vert
G\rangle =B^{\prime }\vert G\rangle =0$. Similar to the normal
phase, we can calculate the photon number variance in the
super-radiate phase as
\begin{eqnarray}
\gamma &=&2f_{1}^{\prime 2}f_{2}^{\prime 2}+2f_{3}^{\prime 2}f_{4}^{\prime
2}+(f_{1}^{\prime }f_{4}^{\prime }+f_{2}^{\prime }f_{3}^{\prime })^{2}
\notag  \label{LEsuper1} \\
&&+\alpha \left[(f_{1}^{\prime }+f_{2}^{\prime })^{2}+(f_{3}^{\prime
}+f_{4}^{\prime })^{2}\right].
\end{eqnarray}

Compared with the case of normal phase, the displacement $\alpha$ of
the photon operator appears in the photon number variance.

\begin{center}
\begin{figure}[th]
\includegraphics[width=8cm]{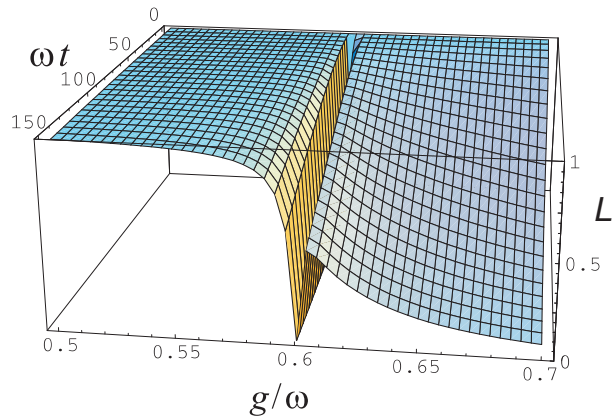}
\caption{(Color online) 3D diagram of the LE plotted as a function
of the time $t$ and the coupling strength $g$ both in the normal
phase (the left panel) and in the super-radiant phase (the right
panel). Here, in unit of $\protect\omega $, $\protect\omega
_{0}=1.44\omega$, $\tilde{ \protect\delta}=g_{s}^{2}/\Delta
_{s}=0.001\omega$ ($\Delta _{s}=0.1\omega$, $g_{s}=0.01\omega$), the
critical point $g_{c}=\protect\sqrt{\protect\omega \protect\omega
_{0}}/2=0.6\omega$, the number of atoms $N=100$.} \label{fig2}
\end{figure}
\end{center}

\section{\label{sec:4}Photon Number Variance for Loschmidt Echo}

We have separately calculated the LE of the photon-dressed atom
ensemble perturbed by an extra atom in two quantum phases: normal
phase and super-radiant phase. Our calculations are based on the
short time approximation, but it can cover the main character of the
QPT of the photon-dressed atomic ensemble induced by the extra atom.
As follows, we illustrate the LE versus the coupling strength $g$
and time $t$ by plotting its three-dimensional (3D) contour.

Figure~\ref{fig2} shows the LE as a function of the time $t$ and the
coupling strength $g$ in the normal and super-radiant phases. It is
obvious that the LE, which is calculated from
Eqs.~(\ref{calculate-b}), (\ref{LEnormal}), and (\ref{LEsuper1}),
will have a sudden change near the critical point. Its decay is
highly enhanced at the critical value $g_{c}$. In the normal phase,
the LE decays rapidly to zero as the enlarged coupling strength $g$
of the photon-dressed atomic ensemble approaches the critical point
$g_{c}$. In the super-radiant phase, similarly, the LE decays faster
as the parameter $g$ decreases to the critical point $g_{c}$. Then
the coherence of the extra atom is very sensitive to the dynamical
perturbation of the photon-dressed atomic ensemble near the critical
point.

Meanwhile, in the vicinity of the critical point, the coherence of the
extra atom decreases to zero sharply with time at fixed point of $g$.
The more nearly the work point $g$ approaches the critical point
$g_{c}$, the sharper the decay of the decoherence of the extra atom is.
During this process, the detected atom evolves from a pure state to a
mixed one. Therefore, we can measure the QPT of the photon-dressed
atomic ensemble by exploring the coherence of the detected atom in the
photon-dressed atomic ensemble.

\begin{figure}[tbp]
\includegraphics[width=8cm]{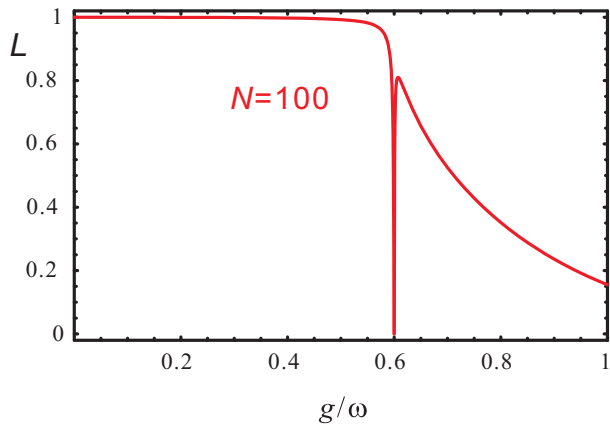}
\caption{(Color online) The cross section of the 3D surface of the
LE in Fig.~\protect\ref{fig2} at $\protect\omega t=100$. For other
parameters see Fig.~\protect\ref{fig2}.} \label{fig3}
\end{figure}

Figure~\ref{fig3} shows the LE at a fixed time ($\omega t=100$) for
the photon-dressed atomic ensemble in both the normal and
super-radiant phases. Contrary to the case of the transverse field
Ising model, the LE in the present system will not approach $1$ when
the coupling strength is much more than the critical point (seen
from Fig.~\ref{fig3}). The reason is that a large displacement
$\sqrt{\alpha }\propto g\sqrt{N}$ appears in the super-radiant phase
and will increase as the coupling strength increases. That means a
small disparity ($\tilde{\delta}a^{\dag }a$) in the initial
Hamiltonian in the super-radiant phase may lead to a large
difference (e.g., the decoherence factor will decay faster) after
period of long-enough time. As pointed out in Ref.~\cite{Emary2003},
the so-called quantum chaos always appears in the super-radiant
phase.

\begin{figure}[tbp]
\includegraphics[width=8cm]{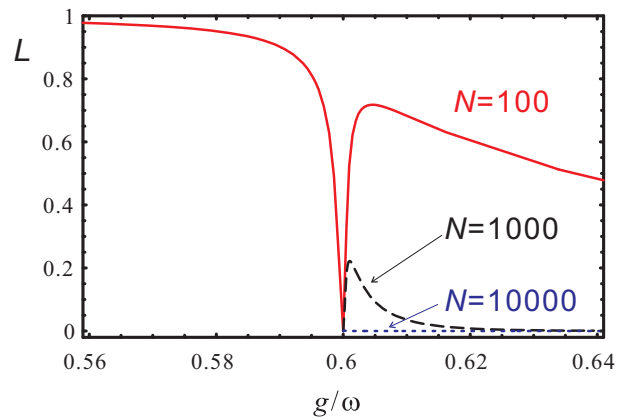}
\caption{(Color online) The LE of the systems for different $N$ at $\protect%
\omega t=100$. In normal phase, the LE is independent of $N$. In
super-radiant phase, $N=100$ (solid line), $1000$ (dashed line), and
$10000$ (dotted line), respectively, from up to bottom. For other
parameters see Fig.~\protect\ref{fig2}.} \label{fig4}
\end{figure}

It follows from Eqs.~(\ref{calculate-b}), (\ref{LEnormal}),
and~(\ref{f}) that, the LE is independent of $N$ in the normal
phase. However, the LE depends on the number of the atoms $N$ in the
super-radiant phase via $\sqrt{\alpha }\propto $ $\sqrt{N}$. In
Fig.~\ref{fig4}, the LE is plotted as a function of the coupling
strength $g$ with $N=100$, $1000$, and $10000$ respectively. It can
be observed from Fig.~\ref{fig4} that the LE line decays faster and
faster in the super-radiant phase as the atom number $N$ increases.
The reason is the same as that mentioned above. The photon number
variance $\gamma $ proportional to the decay rate for the
decoherence of the extra atom increases as $N$ increases via
approximately
\begin{eqnarray}
\gamma \propto \alpha \propto g^{2}N.
\end{eqnarray}
Accordingly, the LE decreases with the form
\begin{eqnarray}
\ln{L} \propto -g^{2}N
\end{eqnarray}
in the super-radiant phase. Thus, as $N\rightarrow \infty $, the
decay of the LE will be strongly enhanced at the critical point.

\section{\label{sec:5}Analog to Cloud Chamber}

Now we can address the similarity of sensitive dynamics between the
present system and the classical cloud chamber. In classical cloud
chamber, when a charged particle (or a dust) flies into the cloud
chamber, which is filled with supersaturated and supercooled water or
alcohol, the water or alcohol vapor will condensate around the flying
charged particle (or a dust) and form a liquid droplet, then a track is
left. During this process, as a result of the sensitivity in response
to the extra particle, the supersaturated vapor staying in the vicinity
of the classical phase transition experiences a classical phase
transition, transiting from vapor to liquid.

In the present investigation, similarly, there exists very sensitive
dynamics of the photon-dressed atomic ensemble when a
far-off-resonant atom goes through the cavity. In view of the Stark
effect, the far-off-resonant atom shifts the frequency of the cavity
field. We assume that the photon-dressed atomic ensemble is
initially prepared in a state near the quantum critical point of the
QPT of the Dicke model. Then the frequency change induced by the
far-off-resonant atom will lead the Dicke model to cross the quantum
critical point, resulting in a sensitive dynamics of the LE. This
quantum effect is similar to the classical phenomenon in the
realistic cloud chamber that the vapor in the cloud chamber will
condensate around the microscopic detected particle after
experiencing the classical phase transition. Therefore, it is
possible to realize the quantum version of the cloud chamber effect
through observing the sensitive change in the LE of the
photon-dressed atomic ensemble.

Here, the enhancement of the decay of LE or its sudden change can be
regarded as an indicator of the one-atom induced QPT to detect the
passage of the atom. This fact properly resembles the cloud chamber
effect. In this analogy, the photon-dressed atomic ensemble, which
can be tuned to the vicinity of the QPT point, behaves as the
supersaturated vapor in the classical cloud chamber, while the
enhancement of the decay of LE just resembles the transition from
vapor to liquid.

Indeed, the LE in our paper is obtained from the decoherence factor
for time evolution of the extra atom, but it actually represents the
``mark" of this atom on the ``cloud chamber" --- the photon-dressed
atomic ensemble. An obvious reason is that the LE only depends on
the parameters of the ``chamber" and, thus, is an intrinsic quantity
of the chamber. Especially, the extra atom can only provide a small
perturbation; thus, the LE is independent of the detected particle.
In most of the references we cite, the LE can be defined without the
detected particle by the chamber. It is only in our own
paper~\cite{Quan2006} where the detected particle is introduced and
it is proved that the decoherence factor of the detected particle is
just the LE of the chamber. Thus, the LE is obviously the mark of
the detected particle left in the chamber.

\section{\label{sec:6}Conclusion with a remark}

In summary, based on the QPT of the Dicke model, we have proposed a
quantum critical model to display the ultra-sensitivity of dynamic
evolution of a QPT system of a photon-dressed atomic ensemble. We
have also pointed out the analog of this one-atom induced QPT to the
cloud chamber based on QPT. Frankly we have to point out that such a
model can not be implemented easily with the generic AMO system,
since the two-photon term could not be simply ignored in the
over-strong coupling limit~\cite{Rzazewski1975}. However, our
present study is still heuristic and the toy model covers the
principle ideas for QPT inducing the cloud chamber-like effect.
Furthermore, with the great development of solid quantum device
physics, the Dicke model may be realized in some solid-state systems
such as the super-conducting quantum circuits and the
nano-mechanical resonators integrated with some qubit array systems.

Finally, we would like to mention a reference~\cite{Carmichael}, in
which an effective Dicke model was derived in a multilevel atomic
ensemble. In this reference, the two-photon term $A^{2}$ may be
safely ignored originally; thus, the modified Dicke model based on
such a practical setup may be used to display the QPT phenomena we
found in this paper.

\begin{acknowledgments}
We would like thank Shuo Yang for helpful discussions. This work was
supported by the National Natural Science Foundation of China with
Grants No. 10935010 and No. 10775048, and the National Fundamental
Research Program of China with Grants No. 2006CB921205 and No.
2007CB925204.
\end{acknowledgments}

\appendix

\section{Coefficients of Bogoliubov transformation}

\subsection{Normal phase}

The coefficients of Bogoliubov transformation in the normal phase are
\begin{eqnarray}
f_{1,2} &=&\frac{1}{2}\frac{\cos \theta }{\sqrt{\omega \omega _{A}}}(\omega
\pm \omega _{A}),  \notag \\
f_{3,4} &=&\frac{1}{2}\frac{\sin \theta }{\sqrt{\omega \omega
_{B}}}(\omega \pm \omega _{B}),\label{f}
\end{eqnarray}
where the rotating angle in the coordinate-momentum representation
$\theta$ is given by
\begin{equation}
\tan 2\theta =\frac{4g\sqrt{\omega \omega _{0}}}{\omega
_{0}^{2}-\omega ^{2}}.
\end{equation}

\subsection{Super-radiant phase}

The coefficients of Bogoliubov transformation in the super-radiant
phase are:
\begin{eqnarray}
f_{1,2}^{\prime } &=&\frac{1}{2}\frac{\cos \theta ^{\prime }}{\sqrt{\omega
\omega _{A}^{\prime }}}(\omega \pm \omega _{A}^{\prime }),  \notag \\
f_{3,4}^{\prime } &=&\frac{1}{2}\frac{\sin \theta ^{\prime }}{\sqrt{\omega
\omega _{B}^{\prime }}}(\omega \pm \omega _{B}^{\prime }),  \label{f-prime}
\end{eqnarray}%
where the analogous rotating angle $\theta ^{\prime }$ is
\begin{equation}
\tan 2\theta ^{\prime }=\frac{2\omega \omega _{0}\mu ^{2}}{\omega
_{0}^{2}-\mu ^{2}\omega ^{2}}.
\end{equation}

\end{document}